\documentclass[aps,prd,twocolumn]{revtex4}
\usepackage{graphicx}
\usepackage{epsfig}
\usepackage{amssymb}
\usepackage{natbib}

\begin{document}

\title{Quasi-Normal Modes of a Schwarzschild White Hole}

\author{Nigel T. Bishop and Amos S. Kubeka}
\affiliation{ Department of Mathematical Sciences, University of
South Africa, P.O. Box 392, Unisa 0003, South Africa}

\begin{abstract}
We investigate perturbations of the Schwarzschild geometry using a
linearization of the Einstein vacuum equations within a Bondi-Sachs,
or null cone, formalism. We develop a numerical
method to calculate the quasi-normal modes, and present results for the
case $\ell=2$. The values obtained are different to those of a Schwarzschild
black hole, and we interpret them as quasi-normal modes of a
Schwarzschild white hole.
\end{abstract}

\maketitle

\section{Introduction}
\label{sec:Introduction}
The theory of a linear perturbation of a black hole was developed
some time ago~\cite{Regge57,Zerilli70,Vishveshwara70,chandrasekhar75};
see also the text-book~\cite{chandrasekhar83} and the review~\cite{Kokkotas99a}.
The essential idea is that the vacuum Einstein equations
are linearized about the Schwarzschild (or Kerr) geometry described
by the usual $(t,r,\theta,\phi)$ coordinates. Then a standard separation
of variables ansatz is applied, with metric quantities behaving as an
unknown function of $r$ $\times Y_{\ell m}(\theta,\phi) \exp(i\nu t)$
(actually, the angular dependence is somewhat more complicated, and the
technical details can be found in the literature). There results an
ordinary differential equation in $r$, and the
quasi-normal modes are obtained by finding the special values of $\nu$
for which solutions exist
that satisfy appropriate boundary conditions
in the neighbouhood of the event horizon, and of infinity.

Quasi-normal mode theory has become a cornerstone of modern general
relativity theory. They have been seen in numerical relativity
simulations of binary black hole coalescence. And, while
not yet actually observed, it is strongly expected that they will
be measured by the LIGO collaboration, and certainly by LISA,
yielding precise information about the parameters describing a
black hole from some coalescence event.

\begin{figure} [ht]
\centering
\epsfig{width=8cm,file=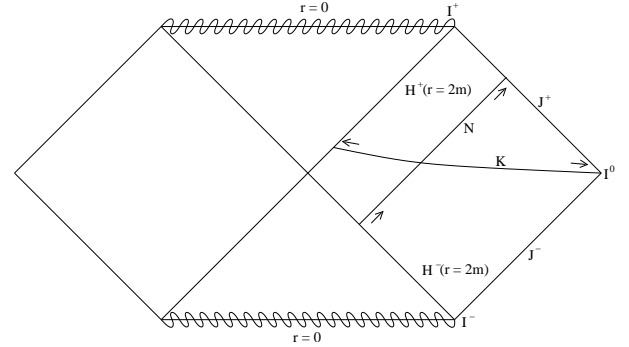}
\caption{Penrose diagram illustrating the differences, in terms of location
and boundary conditions, between the hypersurfaces $K$ and $N$.}
\label{f-penrose}
\end{figure}

In the usual approach to linear perturbations of a black hole,
the linearization is performed using standard Schwarzschild (or
Kerr) coordinates $(t,r,\theta,\phi)$. It is also possible to
perform the linearization using Bondi-Sachs coordinates, which
is a coordinate system based on outgoing null cones. This has
been done in previous work, in order to obtain analytic solutions of
the linearized Einstein equations for the purpose of testing
numerical relativity codes. As with the usual approach, one
ends up with a second order ordinary differential equation involving
$\ell$ and $\nu$ as parameters, Eq.~(\ref{eq:LiEiEQv}). However, when the quasi-normal
modes were calculated for this equation, it was found that they
are not the standard ones.
Different physical problems are considered in the two cases, as
illustrated in the Penrose diagram of Schwarzschild spacetime (Fig.~\ref{f-penrose}).
$K$ is a typical hypersurface used in finding the quasi-normal modes of a
black hole, and the direction of wave propagation at the boundaries
of $K$ is shown by arrows. On the other hand, $N$ is a typical
hypersurface used in finding the quasi-normal modes of Eq.~(\ref{eq:LiEiEQv}).
From the direction of wave propagation on $N$, the resulting quasi-normal
modes can be interpreted as being those of a white hole.

The plan of this paper is as follows. Section~\ref{sec:Linearizedsolution}
summarizes previous work on the Bondi-Sachs metric and linearized solutions
within that framework. Section~\ref{sec:Problem} describes our approach to
calculating the quasi-normal modes, and Sec.~\ref{sec:Results} presents the
results. We end with a Conclusion, Sec.~\ref{sec:Conclusion}.

\section{Background material}

\label{sec:Linearizedsolution}
The Bondi-Sachs formalism uses coordinates $x^{i}=(u,r,x^{A})$ based upon a
family of outgoing null
hypersurfaces. We label the hypersurfaces by $u=$constant, null rays
by $x^{A}$ $(A=2,3)$, and the surface area coordinate by $r$. In this
coordinate system the Bondi-Sachs metric~\cite{Bondi62,Sachs62,Bishop97b} takes the form
\begin{eqnarray}
ds^2=&-&\left[e^{2\beta}\left(1+\frac{W}{r}\right)-r^2h_{AB}U^{A}U^{B}\right]du^2
-2e^{2\beta}dudr\nonumber\\
     &-&2r^2h_{AB}U^{B}dudx^{A}+r^{2}h_{AB}dx^{A}dx^{B},
\label{eq:nulla}
\end{eqnarray}
where $h^{AB}h_{BC}=\delta^{A}_{B}$ and $det(h_{AB})=det(q_{AB})$,
with $q_{AB}$ being a unit sphere metric. We represent $q_{AB}$ by means of a
complex dyad $q_A$. For example, in the
case that the angular coordinates are spherical polar $(\theta,\phi)$,
the dyad takes the form
\begin{equation}
q_A=(1,i\sin\theta).
\end{equation}
For an arbitrary Bondi-Sachs metric,
$h_{AB}$ can be represented by its dyad component
\begin{equation}
J=h_{AB}q^Aq^B/2,
\label{eq:nulle}
\end{equation}
We also introduce the spin-weighted field $U=U^Aq_A,$
as well as the (complex differential) eth operators $\eth$ and $\bar \eth$
~\cite{Gomez97}.
In Schwarzschild space-time, $W=-2M$, $\beta=0$, $U^A=0$ and $J=0$

We use $Z_{\ell m}$, rather than $Y_{\ell m}$, as spherical
harmonic basis functions, where~\cite{Bishop-2005b} the $Z_{\ell m}$ have orthonormal
properties similar to those of the $Y_{\ell m}$, and are real.

We assume the following ansatz, representing a small perturbation of the
Schwarzschild geometry
\begin{eqnarray}
\beta&=&\Re(\beta_{0}(r)e^{i\nu u})Z_{\ell m},\;
        U=\Re(U_{0}(r)e^{i\nu u})\eth Z_{\ell m},\nonumber\\
J&=&\Re(J_{0}(r)e^{i\nu u}) \eth^2 Z_{\ell m},\nonumber \\
W&=&-2M+\Re(w_{0}(r)e^{i\nu u}) Z_{\ell m}.
 \label{eq:nulllebbe}
\end{eqnarray}

Using the above ansatz, Ref.~\cite{Bishop-2005b} constructed the resulting linearized
Einstein vacuum equations. As expected, the angular and time dependence
factored out, and a system of ordinary differential equations (in $r$) was
obtained. As discussed in Ref.~\cite{Bishop-2005b}, the system can be manipulated
to give
\begin{eqnarray}
x^{3}(1-2xM)\frac{d^2J_{2}}{dx^2}
       +2\frac{dJ_{2}}{dx}(2x^2+i{\nu}x-7x^3M) \nonumber\\
       -2(x(\ell^2+\ell-2)/2+8Mx^2+i\nu)J_{2}=0
\label{eq:LiEiEQv}
\end{eqnarray}
where $J_{2}(x)=d^2J_{0}/dx^2$ and $x=1/r$. (Actually, Ref.~\cite{Bishop-2005b} gave
Eq. (\ref{eq:LiEiEQv}) only in the case $\ell=2$, and here we give the formula
for general $\ell$).

\section{Problem specification}
\label{sec:Problem}
We note that Eq. (\ref{eq:LiEiEQv}) has singularities at $x = 0$ and $x = 0.5M$.
The problem is to find values of $\nu$ for which
there exists a solution to Eq. (\ref{eq:LiEiEQv}) that is regular everywhere in
the interval [0, 0.5M]; these values of $\nu$ are the quasi-normal modes. This
is the same situation that is faced when finding the quasi-normal modes of a
black hole. The first solution to this problem was obtained by using series
solutions around the singular points, and a numerical solution of an ordinary
differential equation within the interior of the interval~\cite{chandrasekhar75}.
Subsequently, it was shown \cite{Leaver85} how the theory of 3-term
recurrence relations \cite{Gautschi67} for the series solution about $x=0.5M$
could be used to determine the quasi-normal modes.

It is straightforward to write Eq.~(\ref{eq:LiEiEQv}) with the origin transferred
to $x=0.5M$, and then to evaluate the recurrence relation satisfied by a regular
solution (see Eqs.~(\ref{e-Js}) and (\ref{e-recs}) below). We find a 4-term,
rather than a 3-term,
recurrence relation. While it may be that the quasi-normal modes could be found
by an approach similar to that of~\cite{Leaver85}, this is not a practical option
since there does not seem to be available a well-developed mathematical theory
of 4-term recurrence relations.

Instead, we proceed along the lines used in~\cite{chandrasekhar75}. We construct the
asymptotic series about the essential singularity at $x=0$, and use it to find
a solution to within a specified tolerance at a point $x_0>0$. We then use
this solution as initial data for a numerical solution of Eq.~(\ref{eq:LiEiEQv})
in the range $(x_0,x_c)$ where $x_c<0.5M$; actually, as in~\cite{chandrasekhar75}, we
do not integrate Eq.~(\ref{eq:LiEiEQv}) directly but first convert it to
first-order Ricatti form. Finally, we construct the regular series solution about
$x=0.5M$ and use it to find a solution at $x=x_c$. Then a value of $\nu$ is
a quasi-normal mode if the difference at $x=x_c$ between the regular series
solution and the numerical solution, vanishes.

\subsection{Asymptotic series solution about the essential singularity at $x=0$}

Since the singularity is essential, the resulting series solution has
radius of convergence zero, although it is asymptotic.
We use~\cite{Olver74} to determine rigorous bounds on the error
of approximating the solution by its first $n$ terms.
Note that a series solution $J_2(x)=\sum_{n=1}^\infty a_n x^n$ to
Eq.~(\ref{eq:LiEiEQv}) can be generated by the recurrence relation
\begin{equation}
a_n=-a_{n-1}\frac{n^2+n-6}{2i\nu(n-1)}+a_{n-2}M\frac{2n(n+2)}{2i\nu(n-1)},
\label{e-rec}
\end{equation}
with $a_1=1, a_2=0$.

In order to use the theory developed in~\cite{Olver74}, we must first
transform Eq.~(\ref{eq:LiEiEQv}) to asymptotic form by
\begin{equation}
x \rightarrow z=\frac{1}{x},
\end{equation}
and investigate the solution about the singularity at infinity. We find
\begin{eqnarray}
z^2 (z-2) \frac{d^2 J_2(z)}{dz^2}
-z (2z^2 i\nu +2z -10)\frac{d J_2(z)}{dz}\nonumber \\
-(2z^2 i \nu +4z +16)J_2(z)=0,
\label{e-Jz}
\end{eqnarray}
where we have normalized the scaling of $z$ by setting $M=1$. We evaluate
quantities used in ~\cite{Olver74}:
\begin{eqnarray}
f&=&-\frac{2(z-5+i\nu z^2)}{z(z-2)},\;
g=-\frac{2(2z+8+i\nu z^2)}{z^2(z-2)},\nonumber \\
f_0&=&-2i\nu,\;f_1=-2-4i\nu,\;g_0=0,\nonumber \\
g_1&=&-2i\nu,\;\rho=i\nu,\;\sigma=2+2i\nu.
\end{eqnarray}
Then the solutions can be written as
\begin{equation}
J_{2j}(z) \asymp exp(\lambda_j z) z^{\mu_j} \sum_{s=0}^\infty \frac{a_{s,j}}{z^s}
\end{equation}
where
\begin{equation}
\lambda_1=0,\;\mu_1=-1,\;
\lambda_2=2i\nu,\;\mu_2=3+4i\nu.
\end{equation}

Following~\cite{Olver74}, we let the solution to Eq.~(\ref{e-Jz}) be
\begin{equation}
J_2(z)=L_n (z)+\epsilon_n (z)
\end{equation}
\begin{equation}
\mbox{ where }
L_n(z)=exp(\lambda_1 z) z^{\mu_1} \sum_{s=0}^{n-1} \frac{a_{s,1}}{z^s},
\end{equation}
and define the residual $R_n (z)$ by
\begin{equation}
\frac{d^2 L_n(z)}{dz^2}+f(z)\frac{d L_n(z)}{dz} +g(z) L_n(z)=
\frac{R_n(z)}{z},
\end{equation}
with
\begin{equation}
|R_n(z)|\le \frac{B_n}{z^{n+1}}
\label{e-Rn}
\end{equation}
in some region $|z|>b$ and where $B_n$ is calculable. Ref.~\cite{Olver74}
obtains a bound on $\epsilon_n(z)$ provided the quantity $C(n,b,\nu)$
defined immediately below satisfies $C<1$, where
\begin{equation}
C(n,b,\nu)
=\frac{\beta \sqrt{\pi}\; \Gamma\left(\frac{1}{2}(n+1)+1)\right)}
     {|2i\nu|\Gamma\left(\frac{1}{2}(n+1)+\frac{1}{2})\right)(n+1)},
\end{equation}
where $\beta$ is bounded by
\begin{eqnarray}
\beta &\le&
|4i\nu| +\left|8\frac{1+i\nu}{b-2}\right|+\left|32\frac{1}{b(b-2)}\right|\nonumber \\
        &+&|2i\nu| \left(|2+4i\nu|+\left|2\frac{3-4i\nu}{b-2}\right| \right).
\end{eqnarray}
Given $\nu$ and $b$, we use numerics to determine conditions on $n$ such that
$C<0.99$ and then we bound $\epsilon_n (z)$ by
\begin{equation}
|\epsilon_n (z)|\le \frac{2 B_n}{\beta (1-C(n,b,\nu)|z|^{n+1}}.
\end{equation}

We also need to bound the error $\epsilon^\prime_n (x)$ in using a finite series to
estimate $\frac{dJ_2(x)}{dx}$. Noting that
\begin{equation}
\frac{dJ_2(x)}{dx}=-z^2\frac{dJ_2(z)}{dz},
\end{equation}
the bound on the error is
\begin{equation}
|\epsilon^\prime_n (x)|\le \frac{2 |i\nu|B_n}{\beta (1-C(n,b,\nu)|z|^{n-1}}.
\end{equation}

\subsubsection{Numerical implementation}
We have written Matlab code that takes as input $\nu$ and $b$, and then finds
$\beta$ and the lowest value of $n$ such that $C<0.99$. Then the code finds the
maximum of the absolute values of $\epsilon_n(b)$ and
$\epsilon^\prime_n (x=1/b)$. A bisection method program takes $\nu$ as input and
refines $b$ until the absolute value of the maximum error is in the range
$(0.5,1)\times$ machine precision (about $2\times 10^{-16}$).
The code returns the values of $1/b$ and $L_n(b)/L^\prime_n(x=1/b)$.

\subsection{Numerical integration of Eq.~(\ref{eq:LiEiEQv})}

The first step is to transform Eq.~(\ref{eq:LiEiEQv}) into first-order
Ricatti form. Defining a new dependent variable $v(x)$ by
\begin{equation}
J_2(x) \rightarrow v(x)=\frac{1}{J_2(x)}\frac{dJ_2(x)}{dx},
\label{e-vx}
\end{equation}
we obtain
\begin{eqnarray}
&&x^{3}(1-2x)\left(\frac{dv}{dx}+v^2\right)\nonumber \\
&+&2x(2x+i{\nu}-7x^2)v -2(2x+8x^2+i\nu)=0.
\label{eq:nm-v}
\end{eqnarray}

The numerical integration of Eq.~(\ref{eq:nm-v}) near the singularity
at $x=0$ can be tricky, because we need the result to be as accurate as
possible. We found that a fourth order Runge-Kutta scheme (ode45 in Matlab)
performed better than the stiff schemes, provided stringent tolerance conditions
were used (specifically, RelTol = $10^{-12}$, AbsTol = $10^{-12}$,
MaxStep = $2\times 10^{-6}$). Under these conditions each integration to $x_c$
(=0.25) takes of order 100s.

\subsection{Series solution about the regular singularity at $x=0.5M$}

We first make the transformation
\begin{equation}
x\rightarrow s=1-2x
\label{e-x2s}
\end{equation}
to Eq.~(\ref{eq:LiEiEQv}) and obtain
\begin{eqnarray}
&s&(1-s)^3\frac{d^2J_2(s)}{ds^2}
-(1-s)(4i\nu-3+10s-7s^2)\frac{dJ_2(s)}{ds}\nonumber \\
&-&4(i\nu+3-5s+2s^2)J_2(s)=0.
\label{e-Js}
\end{eqnarray}
This equation has a series solution $\sum_0^\infty a_n s^n$ that
satisfies the recurrence relation
\begin{eqnarray}
&a_0&= 1,\; a_1=4\frac{3+i\nu}{3-4i\nu},\;
    a_2=\frac{15(4+3i\nu)}{2(1-i\nu)(3-4i\nu)}\nonumber \\
&a_n &= a_{n-1}\frac{4ni\nu-8i\nu-5-3n^2-4n}{n(4i\nu-n-2)} \nonumber \\
        &+&a_{n-2}\frac{4+3n^2+2n}{n(4i\nu-n-2)}
        +a_{n-3}\frac{(1-n)(1+n)}{n(4i\nu-n-2)}.
\label{e-recs}
\end{eqnarray}

The radius of convergence of the above series is $s<1$, and, given $\nu$,
the numerical evaluation of the coefficients, and then of the series, is
straightforward. Using $x_c=0.25$ means that we need to evaluate the series
at $s=0.5$. We terminate summation of the series at the first term smaller
than $10^{-18}$ (typically, about 60 terms), and thus expect the result to be
accurate to within machine precision (about $2\times 10^{-16}$).

\section{Results}
 \label{sec:Results}

We have written a Matlab program that, given a value of $\nu$, first uses the
asymptotic series to find the value $v_0$ of $v(x)$ (as defined in Eq.~(\ref{e-vx}))
at $x=x_0=1/b$, and then integrates numerically Eq.~(\ref{eq:nm-v}) between $x_0$ and
$x_c=0.25$, obtaining a complex number $v_+=v(x_c)$; and secondly uses the regular
series about $x=0.5$ to find $v_-=v(x_c)$. Defining
\begin{equation}
g_\nu=v_+ - v_-,
\end{equation}
the quasi-normal modes are those values of $\nu$ such that $g_\nu$ is
indistinguishable from zero.

\begin{figure} [ht]
\centering
\epsfig{width=9cm,file=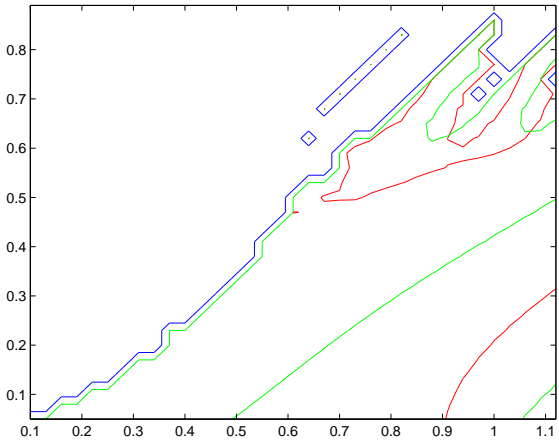}
\caption{Contour plot in the complex plane of $\nu$ showing the contours where
$\Re(g_\nu)=0$ (red) and $\Im(g_\nu)=0$ (green).}
\label{f-cp}
\end{figure}

We calculated $g_\nu$ for values of $\nu$ in the range $\nu=a+ib$,
$0.1\le a\le 1.07$, $0.05\le b\le 0.89$, in increments of 0.03.
The results are shown in the contour plot in Fig~\ref{f-cp}.
The green line is the zero contour of $\Im(g_\nu)$, the red line is the zero
contour of $\Re(g_\nu)$, and the blue line is the boundary of a region where the
computation is probably unreliable (because the computed curve oscillates, indicating
that a smaller step-length is required). Clearly, the quasi-normal modes lie at the
intersection of a red and a green line, and from the plot we can read off an
estimate for the lowest mode, $\nu=0.9+0.63i$. We then applied
a secant method, obtaining a final estimate for the
lowest quasi-normal mode at
\begin{equation}
\nu=0.883+0.614i.
\label{e-lqn}
\end{equation}
In this case, $x_0=0.036493228795438$, $v_0=0.036838521818950 + 0.000637428772012i$,
$\beta=0.988517790240599$, and 62 terms were used in the asymptotic series.
The contour plot indicates another quasi-normal mode at about
$\nu=1.06+0.63i$, but we did not investigate further.

\begin{figure} [ht]
\centering
\epsfig{width=9cm,file=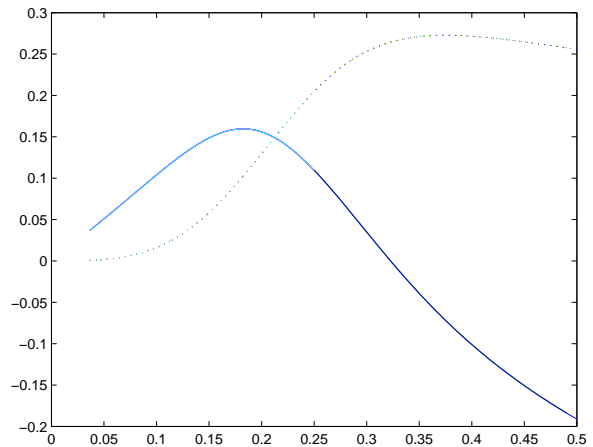}
\caption{The real (solid line) and imaginary (dotted line) parts of $v(x)$
in the quasi-normal mode case $\nu=0.883+0.614i$.}
\label{f-qnm}
\end{figure}

We now use the value in Eq.~(\ref{e-lqn}), and vary the numerical methods so
as to determine the accuracy with which $g_\nu$ has been determined.
In Fig~\ref{f-qnm} the integration between $x_0$ and $x_c$ is carried out
with different values of MaxStep, $2\times 10^{-6}$, $10^{-6}$ and
$5\times 10^{-7}$, and also an error of an amount $(1+i)\times 10^{-15}$
is introduced into the value of $v_0$ at $x_0$ in the case MaxStep =
$2\times 10^{-6}$. Also, numerical integration of Eq.~(\ref{eq:nm-v}) as
well as a series solution is used in the range $(x_c, 0.5)$. The various
curves lie on top of each other and are visually indistinguishable. Taking
all these options into account, the maximum value noted for $g_\nu$ was
$(6.02+5.87i)\times 10^{-4}$. Using intermediate results from the secant
root-finding process to estimate
\begin{equation}
\frac{\partial \nu}{\partial g_\nu}=3.95+0.69i,
\end{equation}
it follows that the possible error in Eq.~(\ref{e-lqn}) is
\begin{equation}
|(3.95+0.69i) \times (6.02+5.87i)\times 10^{-4}| = 0.003,
\end{equation}
so that Eq.~(\ref{e-lqn}) should be amended to read
\begin{equation}
\nu=0.883+0.614i +0.003k
\label{e-lqnc}
\end{equation}
where $k$ is a complex number satisfying $|k|\le 1$.

\begin{figure} [ht]
\centering
\epsfig{width=9cm,file=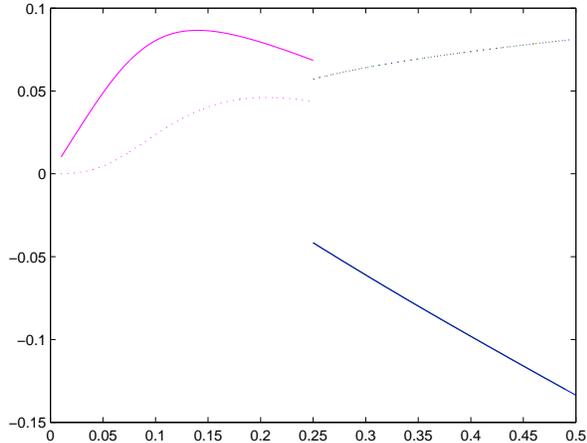}
\caption{The real (solid line) and imaginary (dotted line) parts of $v(x)$
in the case $\nu=0.37367+0.08896i$, indicating that the lowest quasi-normal
mode of a Schwarzschild black hole is not a quasi-normal mode of
Eq.~(\ref{eq:LiEiEQv}).}
\label{f-chqnm}
\end{figure}

The lowest quasi-normal mode of a Schwarzschild black hole is at
$\nu=0.37367+0.08896i$. We
have used this value in our code, and obtained Fig~\ref{f-chqnm}; from which it
is clear that this value of $\nu$ is not a quasi-normal mode of
Eq.~(\ref{eq:LiEiEQv}).

\section{Conclusion}
\label{sec:Conclusion}
Using a linearization of the vacuum Einstein equations about the Schwarzschild
geometry, within a Bondi-Sachs framework, we have constructed a
numerical procedure to calculate the quasi-normal modes. The value of
the lowest mode in the case $\ell=2$ is not a quasi-normal mode of a
Schwarzschild black hole, and further the lowest quasi-normal mode of
a Schwarzschild black hole is not a quasi-normal mode of 
Eq.~(\ref{eq:LiEiEQv}). As discussed in the Introduction, this apparent
discrepancy can be avoided by interpreting the quasi-normal modes of
Eq.~(\ref{eq:LiEiEQv}) as being those of a white hole rather than those of
a black hole.

The results obtained depend crucially on the validity of
Eq.~(\ref{eq:LiEiEQv}), and thus it is important to discuss the extent
to which this has been verified.
Eq.~(\ref{eq:LiEiEQv}) was derived in~\cite{Bishop-2005b}, and there has
been no subsequent, independent, derivation. Nevertheless,
Eq.~(\ref{eq:LiEiEQv}) has been subject to some consistency checks, since
ref.~\cite{Bishop-2005b} confirmed that solutions obtained also satisfy
the remaining Einstein equations (the constraint equations). Further, in
the case $M=0$, solutions based on Eq.~(\ref{eq:LiEiEQv}) have been used
as analytic solutions for the testing of numerical relativity codes based
on the Bondi-Sachs metric, and the expected order of convergence was 
observed~\cite{Babiuc:2009,Reisswig:2006}.

The evidence for the existence of black holes is now very strong.
However, the question about the existence of white holes is much
more problematic, since such objects cannot form from regular initial
data, but instead must have been created as part of the creation of the
universe. The present work provides a possible observational
signature of a white hole,
since it is in principle possible for a gravitational wave detector to
extract the parameters of a quasi-normal mode from a gravitational wave signal.

\section*{Acknowledgments}

NTB and ASK would like to thank the National Research Foundation of
South Africa for financial support. NTB thanks
Max-Planck-Institut f\" ur Gravitationsphysik, for hospitality.
We thank E. Rosinger for
discussion and for drawing our attention to ref.~\cite{Olver74}.

\bibliography{aeireferences}
\end{document}